\documentclass[twocolumn,5p]{elsarticle}

\usepackage{comment}
\usepackage{graphicx,rotating,amsmath,amsfonts,amssymb,amsthm}
\newcommand{\PRexp}{2.14 \pm 0.05}

\begin{document}

\title{Super-Heavy Dark Matter -- Towards Predictive Scenarios from Inflation}

\author[NICPB]{Kristjan Kannike}
\address[NICPB]{National Institute of Chemical Physics and Biophysics, R\"avala 10, 10143 Tallinn, Estonia.}

\author[NICPB]{Antonio Racioppi}

\author[NICPB,Tartu]{Martti Raidal}
\address[Tartu]{ Institute of Physics, University of Tartu, W. Ostwaldi 1,  50411 Tartu, Estonia.}

\date{\today}

\begin{abstract}
{A generic prediction of the Coleman-Weinberg inflation is the existence of a heavy particle sector whose interactions with the inflaton,
the lightest state in this sector,  generate the inflaton potential at loop level. For typical interactions the heavy sector may contain
stable states whose relic abundance is generated at the end of inflation by the gravity alone.
This general feature, and the absence of any particle physics signal of dark matter so far, 
motivates us to look for new directions in the dark sector physics, including scenarios in which dark matter is super-heavy. In this article we study the possibility that
the dark matter is even heavier than the inflaton, its existence follows from the inflaton dynamics, and its abundance today is {\it naturally} determined by
the weakness of gravitational interaction. This implies that the super-heavy dark matter scenarios can be tested via the measurements of inflationary parameters
and/or the CMB isocurvature perturbations and non-Gaussianities.
We explicitly work out details of three Coleman-Weinberg inflation scenarios, study the systematics of super-heavy dark matter production in those cases,
and compute which parts of the parameter spaces can be probed by the future CMB measurements.
}
\end{abstract}


\maketitle

\section{Introduction}
\label{sec:Intro}

Cosmological measurements have shown convincingly that most of the cold matter in the Universe is in a non-baryonic form~\cite{Ade:2015xua}.
The currently dominating paradigm describes the dark matter (DM) as a relic density of weakly interacting massive particles (WIMPs)~\cite{Jungman:1995df}.
However, the extensive experimental program set up for WIMP detection in the direct and indirect detection experiments as well as
in the Large Hadron Collider (LHC) has given negative or inconclusive results so far.  While the discovery of WIMPs in next generation experiments
is still a viable option, the current experimental status suggests that responsible physicists should begin to seriously consider alternatives to the WIMP paradigm.


Alternative views on the DM  have  already been developed. The most radical of them suggests that the DM interacts only gravitationally, explaining the negative experimental results.
For example, the cold DM could be a manifestation of the gravitational sector itself consisting of massive gravitons of bi-metric gravity~\cite{Babichev:2016hir,Aoki:2016zgp} --
the only known self-consistent, ghost-free extension of General Relativity with massive spin-2 fields~\cite{Hassan:2011zd}.

Somewhat less radically, the DM could consist of
gravitationally produced non-thermal relic of supermassive particles -- the  Super-Heavy Dark Matter
(SHDM)~\cite{Chung:1998zb,Kuzmin:1998uv,Kuzmin:1998kk,Kolb:1998ki,Chung:1999ve,Kuzmin:1999zk,Chung:2001cb}.
Indeed, the gravitational production of particles during inflation is the only experimentally verified DM production mechanism as the observed Cosmic Microwave Bacground (CMB)
fluctuations have exactly the same origin. At the end of inflation a fraction of fluctuations are not stretched beyond the horizon but remain as particles because
the inflation slows down. The weakness of gravitational interaction naturally explains the tiny initial abundance of those particles.
In order such an abundance to be cosmologically relevant today, those particles must be superheavy. As a result, the SHDM is the most natural candidate of DM,
even more natural than the WIMP since  the WIMP miracle has no experimental verification yet.

However, the main conceptual problem of the SHDM paradigm has been the lack of any  reason supported  by experimental data why such a particle must exist.
We know the existence of  two scales in Nature, the electroweak scale that sets masses of the standard model (SM) particles and the Planck scale that
determines the strength of gravity.  The existence of  particles at the intermediate scale, ${\cal O}(10^{12}-10^{14})$~GeV, requires additional assumptions.
In the original works~\cite{Chang:1996rf,Chang:1996vw,Faraggi:1999iu,Coriano:2001mg,Chung:1998zb,Kuzmin:1998uv,Kuzmin:1998kk,Kolb:1998ki,Chung:1999ve,Kuzmin:1999zk,Chung:2001cb} as well as in later
papers on the same topic~\cite{Kolb:2007vd,Hikage:2008sk,Chung:2011xd,Chung:2011ck,Chung:2013rda,Aloisio:2015lva}
the existence of such a superheavy particle was connected to the existence of supersymmetry breaking theories, string inspired models or to Kaluza-Klein theories of extra dimensions.
At that time it was commonly believed that naturalness of the
electroweak scale guarantees the existence of new physics. Today this point of view has changed.
The discovery of the Higgs boson~\cite{Chatrchyan:2012xdj,Aad:2012tfa} and the lack of any signal of new physics at the LHC and in any other experiment have challenged
the previously dominating paradigm of naturalness. In the light of those experimental results,
various new theoretical frameworks have been developed to explain  the co-existence and the origin of the largely separated mass scales observed in
Nature~\cite{Heikinheimo:2013fta,Farina:2013mla,Gabrielli:2013hma,deGouvea:2014xba}.
Those attempts involve, in a way or another, the Bardeen's idea of classical scale invariance of the SM~\cite{Bardeen:1995kv},
and the Coleman-Weinberg's idea of loop-level dimensional transmutation~\cite{Coleman:1973jx}.

The aim of this work is to propose that the existence of SHDM may follow from the Coleman-Weinberg  inflation. Indeed, in this framework the
inflaton potential must have been generated at loop level due to the inflaton couplings to new particles in such a way that the experimentally measured
scalar spectral index $n_s$ and the tensor-to-scalar ratio $r$ are predicted correctly. This requirement necessarily introduces entirely new particle sector which may
consist of new singlet scalars and/or fermions (depending on model building details), and which is very weakly coupled to the SM.
Because the inflaton develops a vacuum expectation value (VEV) via dimensional transmutation, it generates the mass scale of the new sector.
Being the pseudo-Goldstone boson of classical scale invariance, the inflaton's mass is suppressed by its $\beta$-function implying that the inflaton itself
is the lightest particle of the sector. The new sector can naturally accommodate stable particles that are viable candidates of the SHDM. For example the lightest fermion of the sector is automatically stable. Similarly, a CP symmetry can act like a $\mathbb{Z}_{2}$ discrete group that stabilises the lightest CP-odd scalar. Therefore, the generic prediction of the proposed scenario is that the SHDM is {\it heavier} than the inflaton.

This approach opens several new perspectives in studies of SHDM. First, it provides a consistent framework for the SHDM model building.
Second, it allows for experimental tests of the SHDM models as the measurable inflationary parameters as well as the reheating temperature
of the Universe are predicted in terms of few model parameters related to the SHDM. Third, also the CMB isocurvature perturbations  and non-Gaussianities
that are always accompanied with the SHDM production -- and that allow for additional tests of the scenario -- are computable. In general, they are predicted to be small
due to the heavier-than-inflaton SHDM.

To demonstrate all those points we study in detail three models of Coleman-Weinberg inflation~\cite{Kannike:2014mia,Kannike:2015kda,Kannike:2015apa}
which all predict $(n_s,r)$ consistently with the present Planck/BICEP2/Keck Array measurements~\cite{Ade:2015xua,Ade:2015tva,Ade:2015fwj,Ade:2015lrj,BICEP2new}.
 We show that all three models are viable candidate models to accommodate the SHDM
which can be largely discriminated from each other with precise measurements of $(n_s,r)$. At the same time, we show that only one model,
the one in which the reheating temperature of the Universe is a free parameter, can generate the isocurvature perturbations at observable level.
We hope that this work represents a step towards renewing interest to the SHDM as the viable and experimentally testable candidate of DM.

This paper is organised as follows.  In Section \ref{sec:CW_and_DM} we discuss model independent features of Coleman-Weinberg inflation and the corresponding phenomenological predictions. In Section~\ref{sec:SHDM} we work out details of three models demonstrating explicitly the connection between Coleman-Weinberg inflation and SHDM. In Section~\ref{subsec:Indirect_probes} we study the experimental constraints and the CMB signatures of this scenario. We present our conclusions in
Section~\ref{sec:Conclusions}.

\section{Generalised Coleman-Weinberg Inflation} \label{sec:CW_and_DM}

We start with covering the essential features of the Coleman-Weinberg inflation in a model independent fashion. Concrete example models will be worked out in next Section.

The concept of Coleman-Weinberg inflation goes back to the very first papers on inflation~\cite{Linde:1981mu,Albrecht:1982wi,Linde:1982zj,Ellis:1982ws,Ellis:1982dg}.
For example,  the Coleman-Weinberg inflation has been studied in the context of grand unified theories
\cite{Langbein:1993ym,GonzalezDiaz:1986bu,Yokoyama:1998rw,Rehman:2008qs} and in the $U(1)_{B-L}$ extension of the
SM~\cite{Barenboim:2013wra,Okada:2013vxa}.  In those models the dynamics leading to dimensional transmutation is
assumed to be new gauge interaction beyond the SM.
However, the dimensional transmutation  occurs most simply via running of scalar quartic coupling $\lambda(\mu)\phi^4$ due to its  coupling  to another scalar field~\cite{Hempfling:1996ht,Gabrielli:2013hma}, generating a non-trivial inflaton potential as demonstrated in Ref.~\cite{Kannike:2014mia}.
The Coleman-Weinberg inflation in the presence of a non-minimal coupling to gravity, $\xi$,
has previously been studied by several authors ~\cite{Okada:2011en,Panotopoulos:2014hwa,Okada:2014lxa,Marzola:2015xbh}.
Alternatively, in \cite{Kannike:2015kda,Cerioni:2009kn,Rinaldi:2015yoa,Barrie:2016rnv} also the Planck scale is dynamically generated via the inflaton vacuum expectation value and its non-minimal coupling to gravity. Coleman-Weinberg inflation models without any explicit mass scale have been presented in~\cite{Kannike:2015apa}.

All such constructions, however, share the same idea of the tree-level quartic inflaton potential $ V^J(\phi) = \frac{1}{4} \lambda_\phi \phi^4$ in the Jordan frame and loop-level dimensional transmutation. Therefore we group them under a common label of Generalised Coleman-Weinberg (GCW) inflation.\footnote{For a different use of the Coleman-Weinberg mechanism and its connection to composite Higgs models, see \cite{Croon:2015fza,Croon:2015naa}.} The most general Lagrangian for the GCW inflation is given by
\begin{equation}
  \sqrt{- g^{J}} \mathcal{L}^{J} = \sqrt{- g^{J}} \left[\mathcal{L}_R + \mathcal{L}_\phi+ \mathcal{L}^{J}_{\sigma, \psi, A_\mu} + \Lambda^4
   \right],
   \label{eq:Jordan:Lagrangian}
\end{equation}
where
\begin{equation}
\mathcal{L}_R = -\frac{M_{EH}^2 + \xi_\phi \phi^{2}}{2}  R \label{eq:L:R}
\end{equation}
contains the Einstein-Hilbert Lagrangian term and a non-minimal coupling, $\xi_\phi$, between the inflaton $\phi$ and the Ricci scalar $R$,
\begin{equation}
\mathcal{L}_\phi = \frac{(\partial \phi)^{2}}{2} - V^J(\phi)
\end{equation}
is the inflaton Lagrangian, $\mathcal{L}^{J}_{\sigma, \psi, A_\mu}$ is the extra matter Lagrangian that induces the one-loop corrections to $V^J(\phi)$, and $\Lambda$ is a cosmological constant term.
All the possible realisations of GCW inflation share two phenomenological constraints. First of all, eq.~\eqref{eq:L:R} should reproduce the observed gravitational coupling, {\it i.e.,}
\begin{equation}
 M_{EH}^2 + \xi_\phi v_\phi^{2} = \bar M_P^2 ,
\end{equation}
where $\bar M_P$ is the reduced Plank mass.
Secondly, in order to avoid problems related to the eternal inflation \cite{Guth:2007ng}, we must assume that
\begin{equation}
V(v_\phi)_{\rm eff}=0 , \label{eq:no:eternal}
\end{equation}
where $V(\phi)_{\rm eff}$ is the one-loop effective potential of inflation including the contribution from the cosmological constant $\Lambda$.
The GCW inflation requires that the inflaton develops a dynamical VEV different from zero.
It can be easily shown in a model-independent way that this is possible when the following condition is satisfied,
\begin{equation}
 \frac{1}{4} \beta_{\lambda_\phi} (v_\phi) + \lambda_\phi (v_\phi) =0, \label{eq:min:eq:V:eff}
\end{equation}
where $\beta_{\lambda_\phi}$ is the $\beta$-function of the running quartic inflaton self-coupling $\lambda_\phi$, defined as $\beta_{\lambda_\phi} = d\lambda_\phi/d(\ln \mu)$ ($\mu$ is the renormalisation scale). Therefore eq.~(\ref{eq:min:eq:V:eff}) becomes a boundary condition on the solution of the RGEs.
As shown in \cite{Kannike:2015kda}, eq.~\eqref{eq:min:eq:V:eff} allows only two types of solutions,
\begin{equation}
 \beta_{\lambda_\phi} (v_\phi) > 0, \ \lambda_\phi (v_\phi)<0 \label{eq:RGE:bound:cond:1} ,
 \end{equation}
 or
\begin{equation}
 \beta_{\lambda_\phi} (v_\phi) = \lambda_\phi (v_\phi)= 0 \label{eq:RGE:bound:cond:2}.
\end{equation}
 The two configurations imply different physics, as we discuss in the following.

In the minimal models~\cite{Kannike:2014mia,Kannike:2015kda,Kannike:2015apa} the generation of Coleman-Weinberg potential for the inflaton $\phi$ requires at the least the presence of one extra scalar $\sigma$. In the rest of this Section we assume such a configuration. The extension to a larger number of similar fields -- which may be needed to accommodate SHDM -- is  straightforward.
According to which model setup we choose, the solution of eq.~(\ref{eq:no:eternal}) requires the tuning of one extra parameter.
The configuration of eq.~(\ref{eq:RGE:bound:cond:1}) requires fine-tuning of the cosmological constant parameter $\Lambda$ so that
\begin{equation}
 \frac{1}{4}\lambda_\phi (v_\phi) v_\phi^4 + \Lambda^4 = 0 .  \label{eq:min:eq:Lambda}
\end{equation}
Alternatively, the configuration of eq.~(\ref{eq:RGE:bound:cond:2}) requires also the presence of an extra fermion $\psi$ and the fine-tuning of  $y_\phi$, its Yukawa coupling to $\phi$, given by the solution of $\beta_{\lambda_\phi} (v_\phi)= 0$. This last configuration automatically implies a vanishing cosmological constant $\Lambda=0$, since we have to enforce the condition $\lambda_\phi (v_\phi)= 0$  to obtain a \emph{global} minimum.
As we do not know the dynamical solution to the cosmological constant problem, at the present stage we are forced to accept the fine-tuning as a solution.
Therefore, according to the choice of the configuration, either (\ref{eq:RGE:bound:cond:1}) or (\ref{eq:RGE:bound:cond:2}), we can construct different models and get different inflationary predictions.

\begin{figure}[t!]
\centering
 \includegraphics[width=0.47\textwidth]{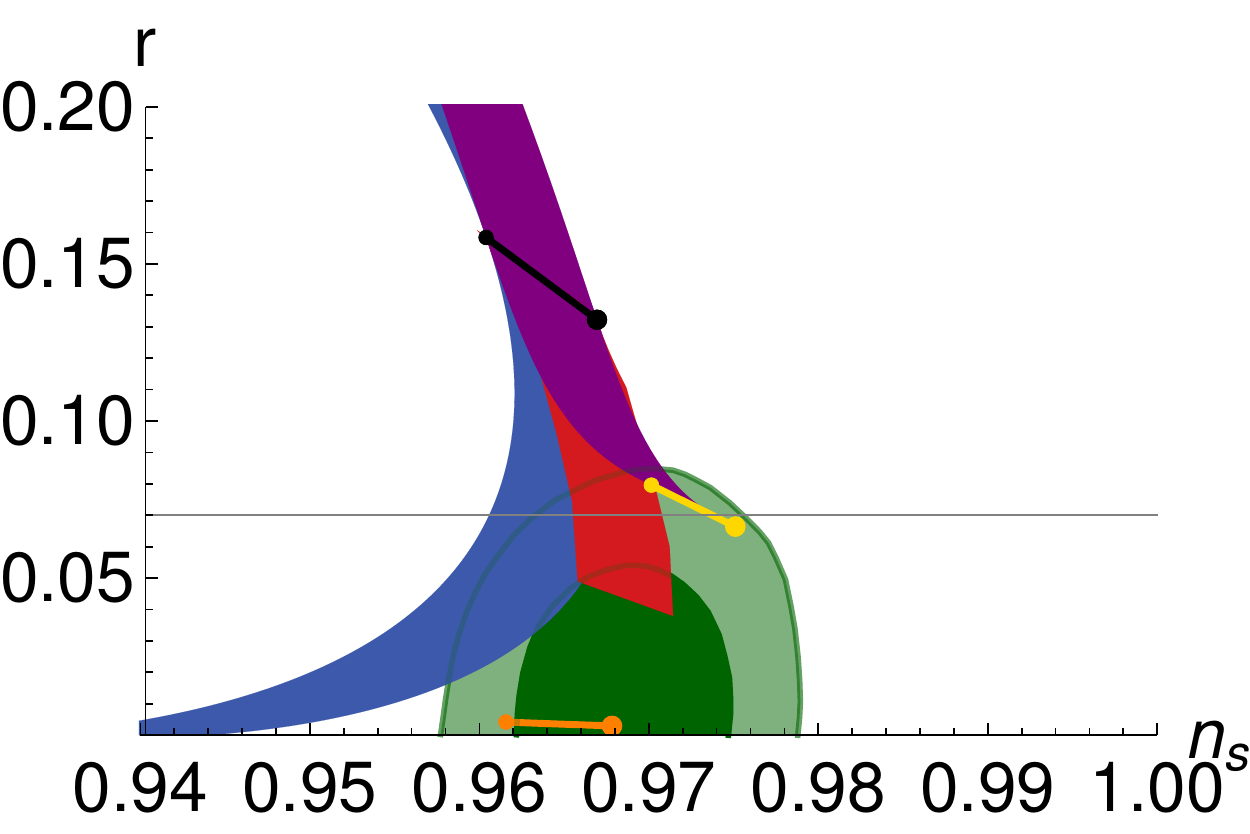}
 \caption{Predictions for tensor-to-scalar ratio $r$ for $N \in [50,60]$ $e$-folds as functions of $n_s$ for three GCW models. The blue region represents the example scenario A, the purple region the scenario B, and the red region the scenario C. For reference the predictions of quadratic, linear and Starobinsky inflation are respectively given by the black, yellow and orange lines. The light green areas present the 1,2$\sigma$ best fits to BICEP2/Keck data \cite{BICEP2new}. The grey line represents $r=0.07$.}
 \label{fig:r:vs:ns:CW}
\end{figure}

In order to study the SHDM in different Coleman-Weinberg inflation scenarios we choose to work with three of them which have different parametric dependence.
Those scenarios are based on different assumptions and, consequently, have different predictions in the $(n_s,r)$ plane and different asymptotic behaviours.
Therefore, future precise experiments measuring the spectral index and the tensor-to-scalar ratio should be able to discriminate between those scenarios.
We present in Fig.~\ref{fig:r:vs:ns:CW}  the predictions for the tensor-to-scalar ratio $r$ for $N \in [50,60]$ $e$-folds of inflation as a function of $n_s$
for three generic Coleman-Weinberg scenarios characterised by the following behaviour.
\begin{itemize}

\item[A.] We assume minimal coupling of the inflaton $\phi$ to gravity, that is $\xi_\phi=0$, and the running of inflaton quartic is dominated by its portal coupling to another scalar
$\sigma$ predicting $\lambda_\phi (\phi) \simeq \beta_{\lambda_\phi} \ln\frac{\phi}{\mu}$.
The asymptotic limit of $(n_s,r)$ in this model is outside the experimentally favoured region. Reheating in this model requires an additional
inflaton coupling to new fermions $\psi_i$.
The new fermions can be identified, for example, with the right-handed neutrinos $N_i$.
Therefore, the reheating temperature $T_{\rm RH}$ depends on the unspecified Yukawa couplings and is a free parameter of the model. This scenario\footnote{This scenario is just the minimal realization of CW inflation, using a scalar instead of a vector field for generating the inflaton loop corrections, therefore the concerning results can be generalized to others CW inflation model \cite{Langbein:1993ym,GonzalezDiaz:1986bu,Yokoyama:1998rw,Rehman:2008qs,Barenboim:2013wra,Okada:2013vxa}.} was studied in \cite{Kannike:2014mia} and is represented in Fig.~\ref{fig:r:vs:ns:CW} by the blue region. 

\item[B.] We assume no explicit Planck mass, $M_{EH}=0$, while the scale of gravity is dynamically generated by the inflaton's  non-minimal coupling to gravity, $\xi_\phi\neq 0$.
As in the previous case,  $\lambda_\phi (\phi)=\beta_{\lambda_\phi} \ln\frac{\phi}{\mu}$, with running dominated by its portal coupling to another scalar
$\sigma$, but the predictions for $(n_s,r)$ are very different.
The asymptotic limit for the latter is the one of linear inflation. In the minimal version of this scenario reheating occurs due to the inflaton decays to the massive
SM bosons, $h, W^{\pm}, Z^0,$ via the non-minimal coupling to gravity\footnote{This is immediate to check in the Einstein frame, where the inflaton coupling to the SM scalars is always generated from the kinetic term  after the conformal transformation from the Jordan frame to the Einstein frame \cite{Kannike:2015kda,Kannike:2015apa}.}.
Thus in the minimal model $T_{\rm RH}$ is not a free parameter but is determined by such  non-minimal coupling.
This scenario\footnote{This scenario and scenario C differ from the other non-minimal CW inflation models \cite{Okada:2011en,Panotopoulos:2014hwa,Okada:2014lxa,Marzola:2015xbh}. Those models introduce  the Planck mass in the Lagrangian by hand, while in this scenario is dynamically generated by the inflaton vev. 
The models described in \cite{Okada:2011en,Panotopoulos:2014hwa,Okada:2014lxa,Marzola:2015xbh} represent another configuration of the parameters space, which should not allow for a correct SHDM relic density. They consider an inflaton mass much smaller than $10^{13}$ GeV, therefore the Planck mass suppression and the exponential factor in eq. (\ref{eq:OmegaX:gravity}) will make the SHDM relic density irrelevant in these models.
Moreover our inflationary results agree with the results of the previous works \cite{Cerioni:2009kn,Rinaldi:2015yoa}.
As shown in details in \cite{Kannike:2015kda,Rinaldi:2015yoa}, for this class of models, the logarithmic loop correction is moving the attractor from Starobinsky to linear inflation.
Finally this scenario differs from scenario C because of the different behaviour of the loop corrections \cite{Kannike:2015kda,Kannike:2015apa}.} was studied in \cite{Kannike:2015kda} and is represented in Fig.~\ref{fig:r:vs:ns:CW} by the purple region.

\item[C.] We assume that no dimensionful parameters exist in the gravity sector,  $M_{EH}=0$,  $\Lambda=0,$ and the vanishing value of the inflaton potential at its minimum
is achieved via tuning the dimensionless couplings of the model, implying $\lambda_\phi (\phi) \simeq \beta_{\lambda_\phi}'(\phi) \ln^2\left(\frac{\phi}{\mu}\right)$, where the derivative is taken with respect to $\ln \mu$. To achieve this, the model must contain at least one new scalar $\sigma$ and a new fermion $\psi$.
For a constant $\beta_{\lambda_\phi}'(\phi)$ this construction reproduces the results of quadratic inflation that is one of the asymptotic limits of the scenario.
For more general  values of $\beta_{\lambda_\phi}'(\phi)$ this scenario predicts a tensor-to-scalar ratio $r$ that approaches the value for Starobinsky inflation~\cite{Starobinsky:1980te}.
With just one inflaton field this limit cannot be reached. However, if additional degrees of freedom are added to the model, for example the $R^2$ term as in  \cite{Kannike:2015apa},
the Starobinsky limit can be achieved. Reheating in this scenario is similar to the one in the scenario B.
This scenario\footnote{Many of the arguments of scenario B apply also for scenario C. However the main difference is that, because of the different quantum correction, scenario C is avoiding the linear attractor and it seems it should fall into the Starobinsky one. The requirement of perturbativity of the theory, however, sets an upper bound in the non-minimal coupling to gravity \cite{Kannike:2015apa}, which implies a lower bound in the tensor-to-scalar ratio $r \gtrsim 0.04$.} 
 was studied in \cite{Kannike:2015apa} and is represented in Fig.~\ref{fig:r:vs:ns:CW} by the red region.
\end{itemize}
The light green areas in Fig.~\ref{fig:r:vs:ns:CW} present the 1, 2$\sigma$ best fit results to the
{BICEP2/Keck} data~\cite{BICEP2new}.
The scenarios A and B are based on the constraint in eq.~(\ref{eq:RGE:bound:cond:1}), while the scenario C is based on the constraint in eq.~(\ref{eq:RGE:bound:cond:2}).
The grey line represents $r=0.07$, which is the average $2 \sigma$ upper bound from~\cite{BICEP2new}, that we use as a reference value for the later discussion about SHDM. The choice for such value is motivated by the $r$ dependence of the power spectrum amplitude of the isocurvature perturbations caused by SHDM: it increases with increasing $r$ value (see eq. (\ref{eq:AdeltaXexp}) or ref. \cite{Hikage:2008sk}). Therefore in order to maximize the effect, we focus on the highest possible $r$ average value, which is 0.07.
We can see that each scenario has a region that fits the experimental constraints but, according to the different realisations, different regions of the $(n_s,r)$ plane are covered.
Depending on the scenario, the SHDM candidates, production  and signatures are affected as we study in the following.

\section{Coleman-Weinberg Inflation and SHDM} \label{sec:SHDM}

The  common feature of the generalised Coleman-Weinberg
inflation is the existence of extra matter content that induces quantum corrections to the tree-level inflaton potential. In the minimal models
studied in this section~\cite{Kannike:2014mia,Kannike:2015kda,Kannike:2015apa}
this matter content is given by one or more real singlet scalars $\sigma$ and, when needed, by singlet fermions $\psi_i$. Some of these particles may be stable and play the role of SHDM candidate.

 There are  two well established mechanisms\footnote{For completeness we also mention freeze-in via gravitational interactions \cite{Garny:2015sjg}. However such a mechanism cannot work in this context because it implies relatively strong (close to the perturbative bound) couplings between the inflaton and SM particles.} for super-heavy particle production in the early Universe that we could use to generate the primordial SHDM abundance:
 \begin{itemize}
 \item preheating via the inflaton oscillations~\cite{Kofman:1994rk,Khlebnikov:1996wr,Khlebnikov:1996zt,Prokopec:1996rr,Greene:1997ge,Chung:1998bt,Giudice:1999fb}
 that is dependent on the coupling of SHDM to the inflaton field;
 \item gravitational production from the vacuum fluctuations at the end of inflation~\cite{Chung:1998zb,Kuzmin:1998uv,Kuzmin:1998kk,Kuzmin:1999zk}, that is
  independent of the coupling of SHDM to the inflaton field.
 \end{itemize}
Because of the Coleman-Weinberg mechanism, the inflaton $\phi$ is essentially the pseudo-Goldstone boson of the dynamically broken classical scale invariance. Therefore it is expected to be lighter\footnote{For a scenario in which the SHDM is lighter than inflaton and is produced in the inflaton decays see~\cite{Farzinnia:2015fka}.}  than $\sigma$ and $\psi$ by a loop factor $\mathcal{O}(10^{-3})$~\cite{Kannike:2014mia,Kannike:2015kda,Kannike:2015apa}. Such a large mass hierarchy comes from the assumption of a minimal matter content, one scalar $\sigma$ (and also the minimal fermion sector $\psi_i$), and the minimisation condition (\ref{eq:min:eq:V:eff}), where $\beta_{\lambda_\phi}$ contains the scalar portal and Yukawa coupling contributions, which are directly connected to the generation of masses for the extra matter fields.
As a consequence of the heavy particle content, for most of the parameter space the minimal  models are compatible with the present CMB non-Gaussianity
constraints~\cite{Chung:2011xd,Aloisio:2015lva,Ade:2015ava}.
However, the dark sector may be much more complicated than the minimal scenarios.
 In the following we will discuss in more detail the SHDM phenomenology scenario per scenario, considering both the above presented production mechanisms.
 In order to compare the different models, we choose a reference value of the tensor-to-scalar ratio which is shared by all the three models $r=0.07$ (see Fig. \ref{fig:r:vs:ns:CW}).


 \subsection{Inflationary Scenario A}
In the case of Coleman-Weinberg  the presence of bosonic degrees of freedom is always required by the stability of the effective potential.
 The minimal realisation of the scenario A contains at least one extra scalar -- besides the inflaton -- that realises the constraint (\ref{eq:RGE:bound:cond:1}).
 For $r=0.07$ the inflaton mass is $m_\phi \approx 10^{13}$~GeV~\cite{Kannike:2014mia}.
 Both the preheating and the gravitational production depend on  the reheating temperature $T_{\rm RH}$ and
 the SHDM-to-inflaton mass ratio ${m_X}/{m_\phi}$,
 where we use $X$ to denote the SHDM candidate, the heavy boson $\sigma$ in this particular case.
 It can be easily checked that in the minimal scenario
 \begin{equation}
 \frac{m_X}{m_\phi} \simeq \frac{2^{3/4} \sqrt{\pi }}{\sqrt[4]{\beta_{\lambda_\phi} }}
 \label{eq:scalar:mass:ratio},
 \end{equation}
 where we assumed that $\lambda_{\phi X}$, the portal coupling between $\phi$ and $X$, dominates the beta function of the self-quartic coupling of the inflaton, $\beta_{\lambda_\phi}$.
The measurement of the amplitude of primordial scalar perturbations \cite{Ade:2015xua,Ade:2015lrj},
 \begin{equation}
A_s \pm \Delta A_s =(\PRexp)\times 10^{-9} , \label{eq:As}
 \end{equation}
 %
fixes $\beta_{\lambda_\phi} \approx 10^{-13,-15}$ so that $ {m_X}/{m_\phi} \approx 10^{3-4}$.
Reheating is achieved via addition of  extra fermions $\psi_i$ with Yukawa couplings $y_{ij}$ small enough as to not spoil the loop potential of the inflaton. The kinematical requirement $m_{\psi_i} < m_\phi/2$ must be satisfied for at least one of the fermions while others may be heavier.
Since $m_\psi = y v_\phi$, such a case can be realised naturally.
This implies an upper bound on the reheating temperature $T_{\rm RH} \lesssim 10^8$ GeV, where the exact limit depends on the exact value of the inflaton mass. With the exception of such a bound, the reheating temperature is a free parameter of the model for all practical purposes.

\subsubsection*{ SHDM production via preheating}

The combination of the large mass ratio ${m_X}/{m_\phi}$, a small portal $\lambda_{\phi X}$ and a low reheating temperature $T_{\rm RH}$ is making the preheating production of scalar $X$ in the minimal model inefficient, resulting in a negligible relic abundance for this SHDM candidate  today~\cite{Chung:1998bt}. Extending the dark sector with more scalars cannot solve the problem because of the requirement of not spoiling the inflaton potential\footnote{In \cite{Chung:1998bt} this problem was solved by assuming a SUSY mechanism that protects inflaton potential against potentially large quantum corrections. Obviously this is not the case in the  minimal model we consider.} by the addition of new particles: an efficient preheating requires relatively large  couplings, which are not compatible with the measurement (\ref{eq:As}). Thus the heavy scalars of this model are not viable SHDM candidates. The same applies if instead we consider a fermionic SHDM candidate. We saw in the previous paragraph that the model contains at least one singlet fermion in order to achieve reheating. However, like the Higgs boson in the SM is coupled to mainly lighter fermions but the top, in the same way the inflaton could be coupled to even more additional fermions, which might be SHDM candidates.
From \cite{Giudice:1999fb} we get the estimate for the abundance of fermionic SHDM produced during preheating,
\begin{equation}
\frac{\rho_X}{\rho} \simeq \frac{y_X^2}{3 \pi^2},
\end{equation}
where $y_X$ is the Yukawa coupling between the fermion SHDM and the inflaton. The requirement that such a Yukawa coupling does not spoil the inflaton potential implies
$y_X^2 \ll \sqrt{16 \pi^2 \beta_{\lambda_{\phi}}} \approx 10^{-5,-6}$ and consequently ${\rho_X}/{\rho} \ll 10^{-6,-7}$, resulting again in an under-abundance of the SHDM. Therefore, we conclude that in this scenario preheating is not a viable mechanism to produce the cosmologically relevant abundance of the SHDM.

\subsubsection*{Gravitational production of SHDM}

\begin{figure}[t]
\centering
 \includegraphics[width=0.48\textwidth]{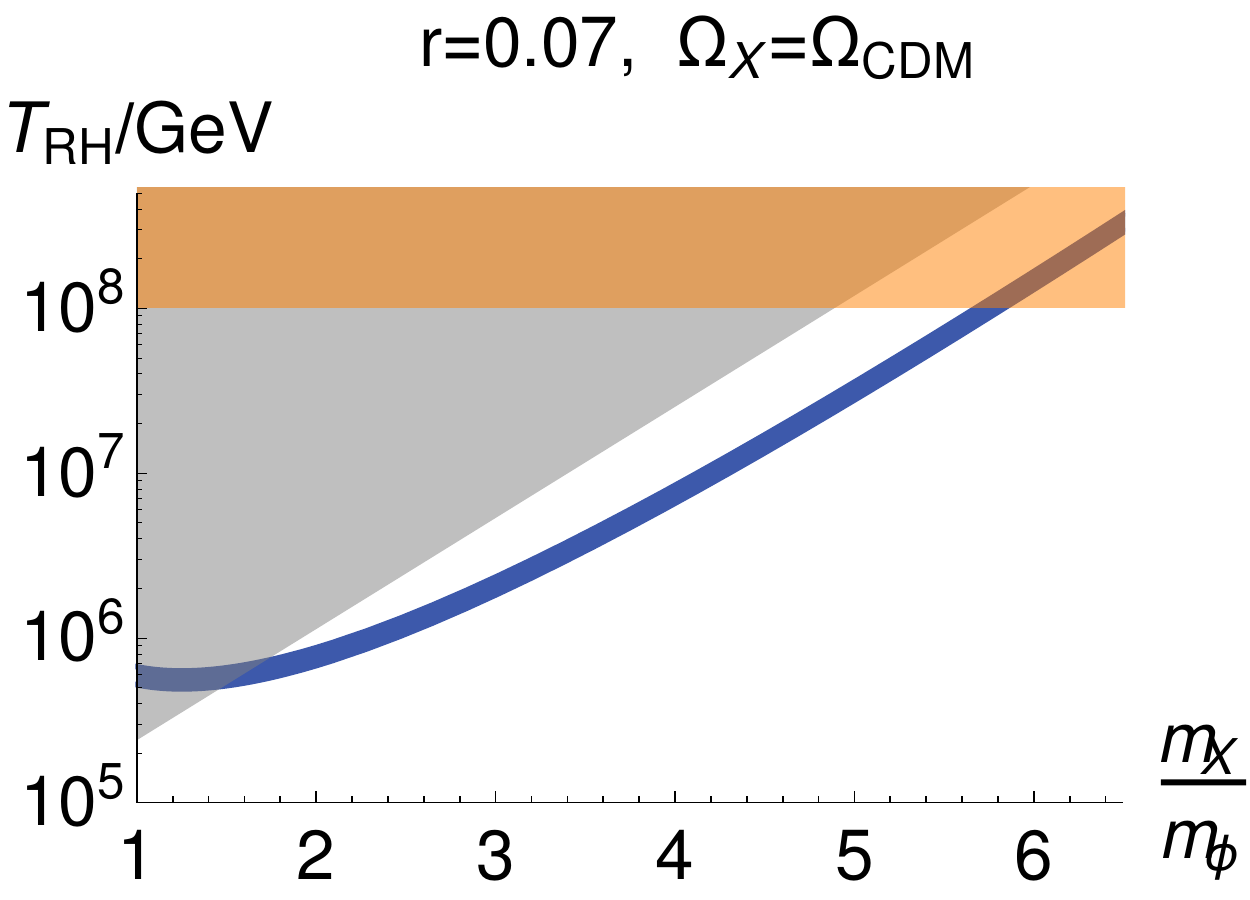}
 \caption{Reheating temperature required to produce all the DM abundance gravitationally, $\Omega_X = \Omega_{\rm CDM}$ \cite{Ade:2015xua}, as a function of the SHDM-to-inflaton mass ratio for the  scenario A. The grey region is  excluded by the constraints on the isocurvature parameter $\alpha$, while the orange region represents unaccessible reheating temperatures in this scenario.}
\label{fig:Omega:vs:mratio:CW}
\end{figure}

Let us start with the scalar SHDM candidate. The relic density of gravitationally produced SHDM today can be approximated as \cite{Kuzmin:1999zk,Chung:2004nh}
\begin{eqnarray}
\hspace{-0.5cm}
\Omega_X (t_0) &\simeq& 10^{-3} \Omega_R \frac{8\pi}{3} \left( \frac{T_{\rm RH}}{T_0} \right)
\left( \frac{m_\phi}{M_P} \right)^2 \left( \frac{m_X}{m_\phi} \right)^{5/2} \times \nonumber\\
&& e^{-2 m_X/m_\phi}, \label{eq:OmegaX:gravity}
\end{eqnarray}
where $\Omega_R \simeq 4 \times 10^{-5}$ is the radiation density today, $T_0 \simeq 2.3 \times 10^{-13}$ GeV is the CMB temperature today and $M_P \simeq 1.22 \times 10^{19}$ GeV is the Planck mass.
We present in Fig.~\ref{fig:Omega:vs:mratio:CW} the required reheating temperature reproducing $\Omega_X = \Omega_{\rm CDM}$ \cite{Ade:2015xua} as a function of the SHDM-to-inflaton mass ratio for the scenario A. The grey region is excluded by the constraints arising from the CMB isocurvature measurements
(to be presented in detail in Section~\ref{subsec:Indirect_probes}), while the orange region represents an upper bound on the reheating temperature in this particular model. This is a simple kinematical bound derived by assuming that the reheating happens through inflaton decay into a pair of fermions and that their mass is generated only by a Yukawa coupling to the inflaton.
We can see that the relic abundance constraint is satisfied for $1.5 \lesssim m_X/m_\phi \lesssim 6$. This ratio is too low to be realised in the minimal model, but is achievable with a more complicated dark sector because such a mass ratio implies a portal coupling that will not spoil the inflaton potential.

As we discussed in the end of Section \ref{sec:CW_and_DM}, we focused our study for $r=0.07$. For smaller $r$ values in the BICEP2 allowed region, the inflaton mass changes by a relatively small amount (maximum around $30\%$) \cite{Kannike:2014mia}, therefore any change in Fig.~\ref{fig:Omega:vs:mratio:CW} will be only minor and impossible to appreciate.

Another possibility is again to consider a richer fermionic sector than the minimal one to achieve reheating (just one singlet fermion). From \cite{Kuzmin:1999zk} we can see that in order to get an appreciable contribution to the relic density for $T_{\rm RH} \approx 10^7$ GeV and $m_\phi \approx 2 \times 10^{13}$ GeV we need $1 \lesssim m_X/m_\phi \lesssim 2$ for the fermion SHDM. Such a mass ratio is allowed because it implies a Yukawa coupling small enough not to spoil the inflaton potential.

\subsection{Inflationary Scenario B}

This scenario involves a non-minimal coupling of the inflaton to gravity. Therefore, it is simpler to look at the theory in the Einstein frame \cite{Kannike:2015kda}. We see that, because of the conformal transformation and the canonical renormalisation of the fields, the inflaton has no portal (or Yukawa) couplings to scalar (fermion) SHDM. Therefore the only available production mechanism of the SHDM is gravitational. Moreover this scenario does not require the presence of any dark fermions since reheating happens via direct decays into
the SM boson pairs, $hh$, $ZZ$ and $WW$. Therefore the reheating temperature in this scenario is determined by the inflaton's non-minimal coupling to gravity.

\begin{figure}[t!]
\centering
 \includegraphics[width=0.47\textwidth]{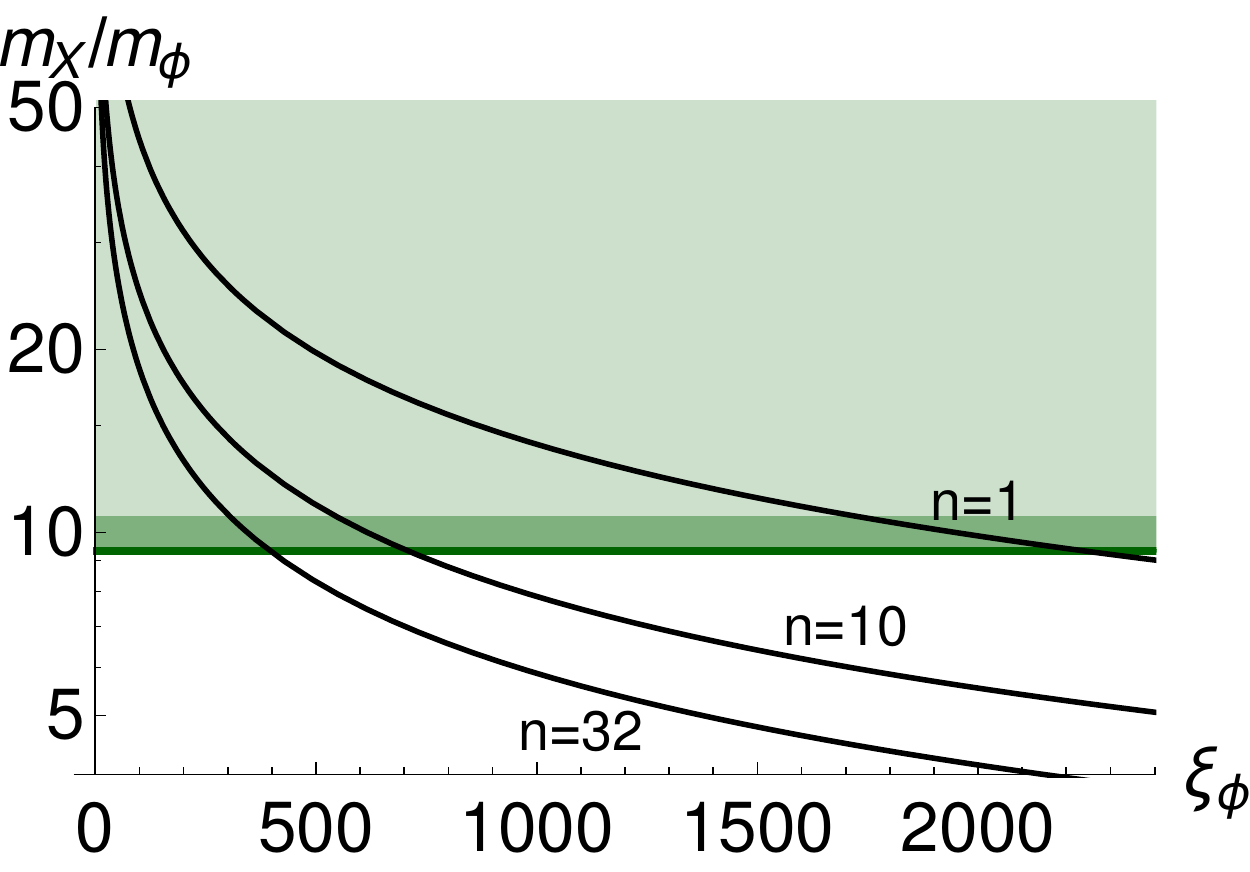}
 \caption{$SO(n)$ SHDM-to-inflaton mass ratio ${m_X}/{m_\phi}$ as a function of the non-minimal coupling to gravity for different $n$ in the scenario B. The narrow dark green region represents the mass range that fits 100\% of the total relic abundance (up to a discrepancy of to two standard devations)~\cite{Ade:2015xua}, while the light and very light green regions represent respectively the mass range that fits more than 10\% and less than 10\% of the total relic abundance.
}
 \label{fig:mratio:vs:xi}
\end{figure}

\subsubsection*{Gravitational production of SHDM}
In this scenario all the physical parameters in eq.~(\ref{eq:OmegaX:gravity}) are depending on the value of the non-minimal coupling $\xi_\phi$ of the inflaton to gravity. To give the model a little bit more flexibility we assume that the scalar $\sigma$ has a $SO(n)$ internal symmetry broken by some dark physics, so that only the lightest state is the SHDM $X$.
Which such an assumption we can generalise eq.~(\ref{eq:scalar:mass:ratio}) to
 \begin{equation}
 \frac{m_X}{m_\phi} \simeq \frac{2^{3/4}}{\sqrt[4]{n \, \beta_{\lambda_\phi} }} \sqrt{\pi} \label{eq:mratio:xiCW}.
 \end{equation}
We recover the minimal model presented in \cite{Kannike:2015kda} for $n=1$.
Then we can use the constraint (\ref{eq:min:eq:Lambda}) to rewrite $\Lambda$ as
 \begin{equation}
\Lambda = \frac{1}{2} v_{\phi } \sqrt[4]{\beta _{\lambda _{\phi }}} \label{eq:Lambda},
 \end{equation}
Given the new BICEP2 data \cite{BICEP2new}, such model is allowed only in the linear limit region, therefore we work in the linear approximation of the Einstein potential, which is \cite{Kannike:2015kda}
 \begin{equation}
V_E(\phi_E) \simeq \frac{\beta_{\lambda _{\phi }} M_P^3}{64 \sqrt{2} \pi ^{3/2} \sqrt{\xi _{\phi }^3 \left(6
   \xi _{\phi }+1\right)}} \phi_E \label{eq:linear},
 \end{equation}
where we used (\ref{eq:Lambda}) and $\phi_E$ is the Einstein frame canonical normalized field value.
Solving the inflationary problem, we can use the constraint (\ref{eq:As}) to fix the normalization of the potential (\ref{eq:linear}) in function of the number of e-folds $N$ getting
 \begin{equation}
 \beta_{\lambda _{\phi }} \simeq 3 \times 10^{-6}  \frac{\sqrt{\xi _{\phi }^3 \left(6 \xi _{\phi}+1\right)}}{(4 N+1)^{3/2}} .
 \end{equation}
Inserting this last equation into eq. (\ref{eq:mratio:xiCW}) we get
 \begin{equation}
 \frac{m_X}{m_\phi} \simeq 72 \frac{\left(4 N+1\right){}^{3/8}}{\sqrt[4]{n} \sqrt[8]{\xi _{\phi }^3
   \left(6 \xi _{\phi }+1\right)}} .
 \end{equation}

For $r=0.07$ we get that $N \simeq 57$ and $m_\phi \simeq 4.5 \times 10^{13}$ GeV.
 Given a value of $\xi_\phi$, we can derive the corresponding $T_{\rm RH}$ \cite{Kannike:2015kda} and, therefore, following the results of \cite{Kuzmin:1999zk},
 the corresponding ${m_X}/{m_\phi}$ as a function of the produced SHDM abundance. Specifically,  we can easily see that in this scenario  $T_{\rm RH} \simeq 6.2 \times 10^{9}$ GeV for $\xi_\phi \gg 1$ \cite{Kannike:2015kda}.

As the result, we plot in Fig.~\ref{fig:mratio:vs:xi} the ratio ${m_X}/{m_\phi}$ a function of the non-minimal coupling $\xi_\phi$ for different values of $n = $ 1, 10 and 32.
The dark green region represents the mass range that fits 100\% of the total relic abundance (up to a discrepancy of to two standard devations)~\cite{Ade:2015xua}, while the light and very light green regions represent respectively the mass range that fits more than 10\% and less than 10\% of the total relic abundance.
We see that in order to produce a cosmologically relevant abundance of the SHDM (more than 10\%) one must have $\xi_\phi \sim \mathcal{O}(10^{2-3})$ depending on the number of bosonic degrees of freedom.

\subsection{Inflationary Scenario C}

Again, it is appropriate to look at this theory in the Einstein frame \cite{Kannike:2015apa}. Because of the conformal transformation and the canonical normalisation of fields the inflaton has no portal (or Yukawa) couplings to the scalar (fermion) candidates of the SHDM. Therefore, the only viable production mechanism of the SHDM is gravitational.

\subsubsection*{Gravitational production}
The minimal model of the scenario C \cite{Kannike:2015apa} requires at least the presence of a scalar $\sigma$ and a fermion $\psi$. In this case the reheating temperature is dominated by decays into SM particles \cite{Kannike:2015apa}, giving $T_{\rm RH} \approx 10^7$ GeV, which is numerically in the same  range as  in the Scenario A. Moreover, also the inflaton mass range is similar, $m_\phi \approx 10^{13}$ GeV. Therefore the prediction for the SHDM mass is essentially the same as in the Scenario A. Because of the large
 ratios  ${m_\sigma}/{m_\phi} ,\, {m_\psi}/{m_\phi} \approx 10^3$~\cite{Kannike:2015apa}, the gravitational production in the minimal model is not efficient. However,  focussing on the fermion sector,  hierarchical Yukawa couplings can easily provide heavy fermions in the appropriate range for the SHDM.  Supposing that the heaviest dark fermion gives the dominant contribution in the solution of eq.~(\ref{eq:RGE:bound:cond:2}), while the lightest is the SHDM that gives the main contribution to the relic density, we get a mass ratio of the same order of the tau-electron mass ratio $m_\tau/m_e$ in the SM. Such a fact may be not a simple coincidence but it could suggest some inner property of the Nature still not yet understood.

\section{CMB and Indirect Probes of the SHDM}
\label{subsec:Indirect_probes}

The main conclusion of  the previous sections is that the gravitationally produced SHDM may be naturally related to the Coleman-Weinberg inflation.
No matter of its nature (scalar or fermion), if this is the correct realisation of the dark sector, we cannot certainly detect any SHDM signal via collider or direct detection experiments. However we can still look for indirect probes of the SHDM like non-Gaussianities in the CMB, isocurvature perturbations or signatures in ultra-high energy cosmic rays. The discussion about the ultra-high energy cosmic rays is given in a detailed way in \cite{Aloisio:2015lva,inprogress}.
In the following we just give a naive discussion of the isocurvature perturbations and non-Gaussianities and their dependence on the  SHDM mass $m_X$. We focus on the scalar case just for simplicity since our scenarios are allowed only for SHDM heavier than the inflaton and non-Gaussianities induced by such a type of fermion SHDM are not properly discussed in the literature yet.

Local non-Gaussianities in the CMB spectrum due to the heavy particles $X$ are parametrised by the effective parameter $f_{NL}^{\rm local}$,
which can be estimated to be \cite{Hikage:2008sk}
\begin{equation}
 f_{NL}^{\rm local} \approx 30 \left( \frac{\alpha}{0.07} \right)^{3/2}.
 \label{fnl}
 \end{equation}
Here
\begin{equation}
 \alpha = \frac{A_{\delta X}}{A_s+A_{\delta X}},
 \end{equation}
where $A_s$ is given by eq.~(\ref{eq:As}), and $A_{\delta X}$ is the power spectrum amplitude of the isocurvature perturbations caused by $X$,
\begin{equation}
\hspace{-0.5cm}
 A_{\delta X} \simeq
  \frac{25 \pi^2 }{96} \frac{M_P^4}{m_X m_{\phi }^3} (A_s r)^2
                     \exp\left({4 \frac{m_X}{m_{\phi }}-\frac{5280 Q}{\pi A_s r}}  \frac{m_X^2}{M_P^2} \right), 
                      \label{eq:AdeltaXexp}
 \end{equation}
where we used the slow-roll approximation, the constraint of the amplitude of primordial scalar perturbations,
and approximated $H_e \approx m_\phi$ and $Q\simeq 2/3$ \cite{Chung:2004nh}.
The current measurement of $f_{NL}^{\rm local} $ by the Planck Collaboration is $\bar f_{NL}^{\rm local} \pm \Delta f_{NL}^{\rm local} = 0.8 \pm 5.0$ \cite{Ade:2015ava}, while the detectability forecast for the large scale structure experiments is $f_{NL}^{\rm local} \sim O(1)$ \cite{Carbone:2008iz,Verde:2009hy}, and in particular $f_{NL}^{\rm local, min} \simeq 3$ according to \cite{Giannantonio:2011ya}.

At the same time the Planck Collaboration is providing constraints directly on the isocurvature parameter
 $\alpha < 0.0019$~\cite{Ade:2015xua}. Since $f_{NL}^{\rm local} $ and $\alpha$ are related via eq.~(\ref{fnl}),
one can easily check that the constraint on $\alpha$ is presently more restrictive than the ones coming from the $f_{NL}$ bounds.
As a result, we find that the bound on the isocurvature parameter $\alpha$  can constrain only the low reheating temperature and
low $m_X/m_\phi$ corner of the parameter space of the gravitationally produced SHDM. For the scenario A this bound is presented
in Fig.~\ref{fig:Omega:vs:mratio:CW} with the grey area. Since the scenario A is the only one in which $T_{\rm RH}$ is a free parameter,
this is the only studied SHDM scenario which is actually constrained by the CMB isocurvature data.
For the remaining scenarios this bound is completely irrelevant.

\begin{figure}[t!]
  \centering
 \includegraphics[width=0.47\textwidth]{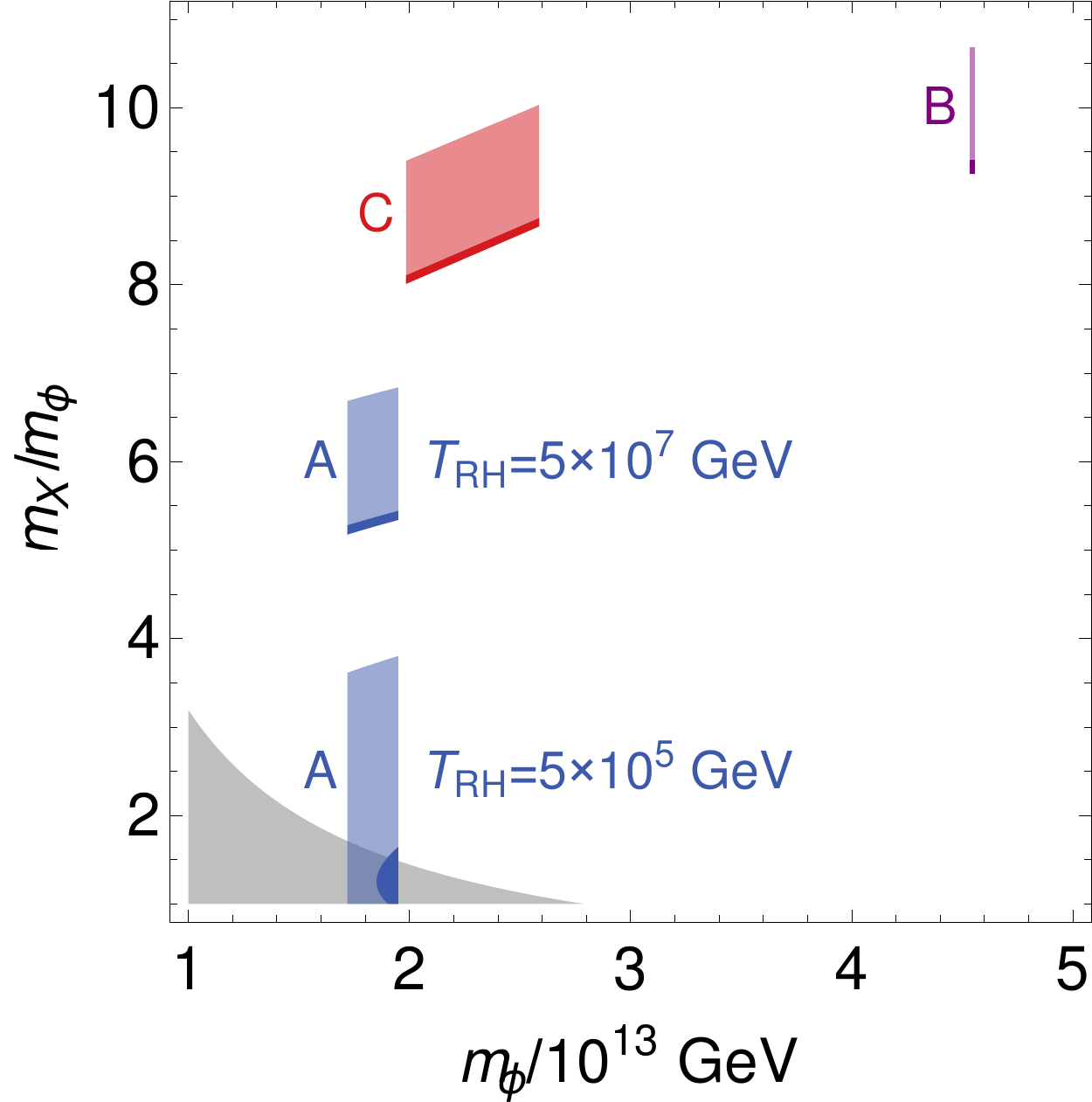}
  \caption{Summary of our results for the scenarios A, B, C represented by the blue/purple/red regions, respectively, on the $(m_\phi, m_X/m_\phi)$ plane for $r=0.07$.
  The dark (light) coloured contours represent the parameter ranges for which the total (10\% of) cosmological DM abundance can be created via gravitational production.
  The grey region is  excluded  by the constraints on the isocurvature parameter $\alpha$.}
  \label{fig:alphaplot}
\end{figure}

Moving to smaller $r$ values, the results remain essentially unaffected. A similar argument of the one of scenario A holds also for scenario C, therefore changing r to smaller values will only slightly change the predicted regions \cite{Kannike:2014mia,Kannike:2015apa}. The most affected region would be the excluded one by constraints on the isocurvature parameter, which will become even smaller (see eqs. (20-21)).

To summarise our findings, we present in Fig.~\ref{fig:alphaplot}  the regions in the $(m_\phi, m_X/m_\phi)$ parameter space for which cosmologically relevant
SHDM abundance can be generated via the gravitational production mechanism.
The blue/purple/red regions represent the scenarios A/B/C, respectively. The dark coloured areas represent $\Omega_X =\Omega_{\rm CDM} \pm 2 \sigma(\Omega_{\rm CDM})$~\cite{Ade:2015xua} while the light coloured areas represent $\Omega_X > 0.1 \Omega_{\rm CDM} $. The inflaton mass range is given by the choice $r=0.07$.
The results for the scenario A are given for two reference values of the reheating temperature, $T_{\rm RH} =5 \times 10^5$ GeV and $T_{\rm RH} =5 \times 10^7$ GeV.
The grey region is the excluded region from the constraints on the isocurvature parameter $\alpha$.
One can see that for low reheating temperatures the scenario A could potentially be testable with more precise CMB isocurvature measurements.
However, in the scenarios B and C the reheating temperature is fixed by the inflaton non-minimal coupling to gravity so that there is no freedom to
choose values of $T_{\rm RH}$. Those scenarios predict too large $m_X/m_\phi$ to be tested by the CMB measurements. Of course, we stress that the results in
Fig.~\ref{fig:alphaplot} apply to the most minimal scenarios studied in this work. Generalising those models with extended particle content would allow
one to construct models that predict isocurvature perturbations within detectable range.

\section{Conclusions and Outlook}
\label{sec:Conclusions}

We have argued that the framework of Coleman-Weinberg inflation naturally motivates the existence of SHDM.
Those  models introduce new heavy particle sector which generates the inflaton potential at loop level
in agreement with the recent Planck/BICEP2/Keck  results~\cite{Ade:2015xua,Ade:2015tva,Ade:2015fwj,Ade:2015lrj,BICEP2new} and may contain heavier-than-inflaton SHDM.
The new sector is very weakly coupled to the SM explaining the coexistence of two vastly separate scales
consistently with the Physical Naturalness principle~\cite{Heikinheimo:2013fta,Farina:2013mla,Gabrielli:2013hma,deGouvea:2014xba}.

To exemplify those general arguments  we studied in details three different  Coleman-Weinberg models of inflation~\cite{Kannike:2014mia,Kannike:2015kda,Kannike:2015apa}
which are different by construction and by particle content and whose predictions cover different regions in the $(n_s, r)$ plane.
For every model we considered both heavy fermions and scalars as the possible candidates for the SHDM. We analysed both preheating and gravitational production mechanisms
of the SHDM, discovering that preheating is never efficient enough to produce the entire DM abundance, leaving the gravitational production as the only viable mechanism.
In all models, the requirement of producing the observed relic abundance  fixes the SHDM mass to be approximately in the range between 1 and 10 inflaton masses,
while the exact SHDM mass depends on the model dependent parameter values.
At the same time,  the requirement of generating observable CMB isocurvature perturbations  sets  an upper bound on the SHDM mass that is approximately 4 inflaton masses.

Our results show that in two out of the three models those two requirements are in conflict with each other, see Fig.~\ref{fig:alphaplot}, leaving only one studied model (scenario A) that
can predict isocurvature perturbations in the potentially observable range. The physical reason for such an outcome is that the first two models are so constrained
that the reheating temperature of the Universe is predicted in terms of other model parameters, rendering to small isocurvature effects. Consequently,
in more general models we expect more observable effects than in the minimal ones.
On the other hand, the most minimal model is presented by scenario B, in which we can make do with just one extra particle besides inflation, the new scalar $\sigma$ which generates the Coleman-Weinberg potential for the inflaton and plays the role of DM.

We would like to finish with a positive note. Although the searches for WIMPs in direct and indirect detection experiments and at colliders may fail,
the alternative DM scenarios like the one presented in this work may provide new ways of experimental tests of the DM of the Universe.
The expected improvement of the measurement precision of the tensor-to-scalar ratio $r$, the CMS isocurvature perturbations and the non-Gaussianity parameters,
or the eventual detection of spectral features in the ultra high-energy cosmic rays or a marked anisotropic flux beyond $10^{20}$~eV (see~\cite{inprogress}),
may confirm the existence of SHDM in the context of Coleman-Weinberg inflationary scenarios. If this will be the case,
our present rough numerical estimates must be refined in terms of concrete models favoured by those experimental results.

\section*{Acknowledgments}

The authors thank Luca Marzola and Federico R. Urban for useful discussions.
This work was supported by  the Estonian Research Council grants IUT23-6, PUT799, PUT1026, and by
EU through the European Regional Development Fund (grant TK133).



\bibliographystyle{JHEP}
\bibliography{references}

\providecommand{\href}[2]{#2}\begingroup\raggedright\begin{thebibliography}{10}

\bibitem{Ade:2015xua}
{\bf Planck} Collaboration, P.~A.~R. Ade et~al., {\it {Planck 2015 results.
  XIII. Cosmological parameters}},  \href{http://arxiv.org/abs/1502.01589}{{\tt
  arXiv:1502.01589}}.

\bibitem{Jungman:1995df}
G.~Jungman, M.~Kamionkowski, and K.~Griest, {\it {Supersymmetric dark matter}},
   {\em Phys. Rept.} {\bf 267} (1996) 195--373,
  [\href{http://arxiv.org/abs/hep-ph/9506380}{{\tt hep-ph/9506380}}].

\bibitem{Babichev:2016hir}
E.~Babichev, L.~Marzola, M.~Raidal, A.~Schmidt-May, F.~Urban, H.~Veerm{\"a}e,
  and M.~von Strauss, {\it {Gravitational Origin of Dark Matter}},
  \href{http://arxiv.org/abs/1604.08564}{{\tt arXiv:1604.08564}}.

\bibitem{Aoki:2016zgp}
K.~Aoki and S.~Mukohyama, {\it {Bigravitons as dark matter and gravitational
  waves}},  \href{http://arxiv.org/abs/1604.06704}{{\tt arXiv:1604.06704}}.

\bibitem{Hassan:2011zd}
S.~F. Hassan and R.~A. Rosen, {\it {Bimetric Gravity from Ghost-free Massive
  Gravity}},  {\em JHEP} {\bf 02} (2012) 126,
  [\href{http://arxiv.org/abs/1109.3515}{{\tt arXiv:1109.3515}}].

\bibitem{Chung:1998zb}
D.~J. Chung, E.~W. Kolb, and A.~Riotto, {\it {Superheavy dark matter}},  {\em
  Phys.Rev.} {\bf D59} (1999) 023501,
  [\href{http://arxiv.org/abs/hep-ph/9802238}{{\tt hep-ph/9802238}}].

\bibitem{Kuzmin:1998uv}
V.~Kuzmin and I.~Tkachev, {\it {Ultrahigh-energy cosmic rays, superheavy long
  living particles, and matter creation after inflation}},  {\em JETP Lett.}
  {\bf 68} (1998) 271--275, [\href{http://arxiv.org/abs/hep-ph/9802304}{{\tt
  hep-ph/9802304}}].

\bibitem{Kuzmin:1998kk}
V.~Kuzmin and I.~Tkachev, {\it {Matter creation via vacuum fluctuations in the
  early universe and observed ultrahigh-energy cosmic ray events}},  {\em
  Phys.Rev.} {\bf D59} (1999) 123006,
  [\href{http://arxiv.org/abs/hep-ph/9809547}{{\tt hep-ph/9809547}}].

\bibitem{Kolb:1998ki}
E.~W. Kolb, D.~J.~H. Chung, and A.~Riotto, {\it {WIMPzillas!}},  in {\em
  {Trends in theoretical physics II. Proceedings, 2nd La Plata Meeting, Buenos
  Aires, Argentina, November 29-December 4, 1998}}, pp.~91--105, 1998.
\newblock \href{http://arxiv.org/abs/hep-ph/9810361}{{\tt hep-ph/9810361}}.
\newblock [,91(1998)].

\bibitem{Chung:1999ve}
D.~J. Chung, E.~W. Kolb, A.~Riotto, and I.~I. Tkachev, {\it {Probing Planckian
  physics: Resonant production of particles during inflation and features in
  the primordial power spectrum}},  {\em Phys.Rev.} {\bf D62} (2000) 043508,
  [\href{http://arxiv.org/abs/hep-ph/9910437}{{\tt hep-ph/9910437}}].

\bibitem{Kuzmin:1999zk}
V.~A. Kuzmin and I.~I. Tkachev, {\it {Ultrahigh-energy cosmic rays and
  inflation relics}},  {\em Phys. Rept.} {\bf 320} (1999) 199--221,
  [\href{http://arxiv.org/abs/hep-ph/9903542}{{\tt hep-ph/9903542}}].

\bibitem{Chung:2001cb}
D.~J. Chung, P.~Crotty, E.~W. Kolb, and A.~Riotto, {\it {On the gravitational
  production of superheavy dark matter}},  {\em Phys.Rev.} {\bf D64} (2001)
  043503, [\href{http://arxiv.org/abs/hep-ph/0104100}{{\tt hep-ph/0104100}}].

\bibitem{Chang:1996rf}
S.~Chang, C.~Coriano, and A.~E. Faraggi, {\it {New dark matter candidates
  motivated from superstring derived unification}},  {\em Phys. Lett.} {\bf
  B397} (1997) 76--80, [\href{http://arxiv.org/abs/hep-ph/9603272}{{\tt
  hep-ph/9603272}}].

\bibitem{Chang:1996vw}
S.~Chang, C.~Coriano, and A.~E. Faraggi, {\it {Stable superstring relics}},
  {\em Nucl. Phys.} {\bf B477} (1996) 65--104,
  [\href{http://arxiv.org/abs/hep-ph/9605325}{{\tt hep-ph/9605325}}].

\bibitem{Faraggi:1999iu}
A.~E. Faraggi, K.~A. Olive, and M.~Pospelov, {\it {Probing the desert with
  ultraenergetic neutrinos from the sun}},  {\em Astropart. Phys.} {\bf 13}
  (2000) 31--43, [\href{http://arxiv.org/abs/hep-ph/9906345}{{\tt
  hep-ph/9906345}}].

\bibitem{Coriano:2001mg}
C.~Coriano, A.~E. Faraggi, and M.~Plumacher, {\it {Stable superstring relics
  and ultrahigh-energy cosmic rays}},  {\em Nucl. Phys.} {\bf B614} (2001)
  233--253, [\href{http://arxiv.org/abs/hep-ph/0107053}{{\tt hep-ph/0107053}}].

\bibitem{Kolb:2007vd}
E.~W. Kolb, A.~Starobinsky, and I.~Tkachev, {\it {Trans-Planckian wimpzillas}},
   {\em JCAP} {\bf 0707} (2007) 005,
  [\href{http://arxiv.org/abs/hep-th/0702143}{{\tt hep-th/0702143}}].

\bibitem{Hikage:2008sk}
C.~Hikage, K.~Koyama, T.~Matsubara, T.~Takahashi, and M.~Yamaguchi, {\it
  {Limits on Isocurvature Perturbations from Non-Gaussianity in WMAP
  Temperature Anisotropy}},  {\em Mon.Not.Roy.Astron.Soc.} {\bf 398} (2009)
  2188--2198, [\href{http://arxiv.org/abs/0812.3500}{{\tt arXiv:0812.3500}}].

\bibitem{Chung:2011xd}
D.~J. Chung and H.~Yoo, {\it {Isocurvature Perturbations and Non-Gaussianity of
  Gravitationally Produced Nonthermal Dark Matter}},  {\em Phys.Rev.} {\bf D87}
  (2013), no.~2 023516, [\href{http://arxiv.org/abs/1110.5931}{{\tt
  arXiv:1110.5931}}].

\bibitem{Chung:2011ck}
D.~J. Chung, L.~L. Everett, H.~Yoo, and P.~Zhou, {\it {Gravitational Fermion
  Production in Inflationary Cosmology}},  {\em Phys.Lett.} {\bf B712} (2012)
  147--154, [\href{http://arxiv.org/abs/1109.2524}{{\tt arXiv:1109.2524}}].

\bibitem{Chung:2013rda}
D.~J. Chung, H.~Yoo, and P.~Zhou, {\it {Fermionic Isocurvature Perturbations}},
   {\em Phys.Rev.} {\bf D91} (2015), no.~4 043516,
  [\href{http://arxiv.org/abs/1306.1966}{{\tt arXiv:1306.1966}}].

\bibitem{Aloisio:2015lva}
R.~Aloisio, S.~Matarrese, and A.~Olinto, {\it {Super Heavy Dark Matter in light
  of BICEP2, Planck and Ultra High Energy Cosmic Rays Observations}},
  \href{http://arxiv.org/abs/1504.01319}{{\tt arXiv:1504.01319}}.

\bibitem{Chatrchyan:2012xdj}
{\bf CMS} Collaboration, S.~Chatrchyan et~al., {\it {Observation of a new boson
  at a mass of 125 GeV with the CMS experiment at the LHC}},  {\em Phys. Lett.}
  {\bf B716} (2012) 30--61, [\href{http://arxiv.org/abs/1207.7235}{{\tt
  arXiv:1207.7235}}].

\bibitem{Aad:2012tfa}
{\bf ATLAS} Collaboration, G.~Aad et~al., {\it {Observation of a new particle
  in the search for the Standard Model Higgs boson with the ATLAS detector at
  the LHC}},  {\em Phys. Lett.} {\bf B716} (2012) 1--29,
  [\href{http://arxiv.org/abs/1207.7214}{{\tt arXiv:1207.7214}}].

\bibitem{Heikinheimo:2013fta}
M.~Heikinheimo, A.~Racioppi, M.~Raidal, C.~Spethmann, and K.~Tuominen, {\it
  {Physical Naturalness and Dynamical Breaking of Classical Scale Invariance}},
   {\em Mod. Phys. Lett.} {\bf A29} (2014) 1450077,
  [\href{http://arxiv.org/abs/1304.7006}{{\tt arXiv:1304.7006}}].

\bibitem{Farina:2013mla}
M.~Farina, D.~Pappadopulo, and A.~Strumia, {\it {A modified naturalness
  principle and its experimental tests}},  {\em JHEP} {\bf 08} (2013) 022,
  [\href{http://arxiv.org/abs/1303.7244}{{\tt arXiv:1303.7244}}].

\bibitem{Gabrielli:2013hma}
E.~Gabrielli, M.~Heikinheimo, K.~Kannike, A.~Racioppi, M.~Raidal, et~al., {\it
  {Towards Completing the Standard Model: Vacuum Stability, EWSB and Dark
  Matter}},  {\em Phys.Rev.} {\bf D89} (2014) 015017,
  [\href{http://arxiv.org/abs/1309.6632}{{\tt arXiv:1309.6632}}].

\bibitem{deGouvea:2014xba}
A.~de~Gouvea, D.~Hernandez, and T.~M.~P. Tait, {\it {Criteria for Natural
  Hierarchies}},  {\em Phys. Rev.} {\bf D89} (2014), no.~11 115005,
  [\href{http://arxiv.org/abs/1402.2658}{{\tt arXiv:1402.2658}}].

\bibitem{Bardeen:1995kv}
W.~A. Bardeen, {\it {On naturalness in the standard model}},  {\em
  FERMILAB-CONF-95-391-T, C95-08-27.3} (1995).

\bibitem{Coleman:1973jx}
S.~R. Coleman and E.~J. Weinberg, {\it {Radiative Corrections as the Origin of
  Spontaneous Symmetry Breaking}},  {\em Phys. Rev.} {\bf D7} (1973)
  1888--1910.

\bibitem{Kannike:2014mia}
K.~Kannike, A.~Racioppi, and M.~Raidal, {\it {Embedding inflation into the
  Standard Model - more evidence for classical scale invariance}},  {\em JHEP}
  {\bf 1406} (2014) 154, [\href{http://arxiv.org/abs/1405.3987}{{\tt
  arXiv:1405.3987}}].

\bibitem{Kannike:2015kda}
K.~Kannike, A.~Racioppi, and M.~Raidal, {\it {Linear inflation from quartic
  potential}},  {\em JHEP} {\bf 01} (2016) 035,
  [\href{http://arxiv.org/abs/1509.05423}{{\tt arXiv:1509.05423}}].

\bibitem{Kannike:2015apa}
K.~Kannike, G.~H{\"u}tsi, L.~Pizza, A.~Racioppi, M.~Raidal, et~al., {\it
  {Dynamically Induced Planck Scale and Inflation}},  {\em JHEP} {\bf 1505}
  (2015) 065, [\href{http://arxiv.org/abs/1502.01334}{{\tt arXiv:1502.01334}}].

\bibitem{Ade:2015tva}
{\bf BICEP2, Planck} Collaboration, P.~A.~R. Ade et~al., {\it {Joint Analysis
  of BICEP2/$Keck  Array$ and $Planck$ Data}},  {\em Phys. Rev. Lett.} {\bf
  114} (2015) 101301, [\href{http://arxiv.org/abs/1502.00612}{{\tt
  arXiv:1502.00612}}].

\bibitem{Ade:2015fwj}
{\bf BICEP2, Keck Array} Collaboration, P.~A.~R. Ade et~al., {\it {BICEP2 /
  Keck Array V: Measurements of B-mode Polarization at Degree Angular Scales
  and 150 GHz by the Keck Array}},  {\em Astrophys. J.} {\bf 811} (2015) 126,
  [\href{http://arxiv.org/abs/1502.00643}{{\tt arXiv:1502.00643}}].

\bibitem{Ade:2015lrj}
{\bf Planck} Collaboration, P.~Ade et~al., {\it {Planck 2015 results. XX.
  Constraints on inflation}},  \href{http://arxiv.org/abs/1502.02114}{{\tt
  arXiv:1502.02114}}.

\bibitem{BICEP2new}
{\bf BICEP2, Keck Array} Collaboration, P.~A.~R. Ade et~al., {\it {Improved
  Constraints on Cosmology and Foregrounds from BICEP2 and Keck Array Cosmic
  Microwave Background Data with Inclusion of 95 GHz Band}},  {\em Phys. Rev.
  Lett.} {\bf 116} (2016) 031302, [\href{http://arxiv.org/abs/1510.09217}{{\tt
  arXiv:1510.09217}}].

\bibitem{Linde:1981mu}
A.~D. Linde, {\it {A New Inflationary Universe Scenario: A Possible Solution of
  the Horizon, Flatness, Homogeneity, Isotropy and Primordial Monopole
  Problems}},  {\em Phys.Lett.} {\bf B108} (1982) 389--393.

\bibitem{Albrecht:1982wi}
A.~Albrecht and P.~J. Steinhardt, {\it {Cosmology for Grand Unified Theories
  with Radiatively Induced Symmetry Breaking}},  {\em Phys.Rev.Lett.} {\bf 48}
  (1982) 1220--1223.

\bibitem{Linde:1982zj}
A.~D. Linde, {\it {Coleman-Weinberg Theory and a New Inflationary Universe
  Scenario}},  {\em Phys.Lett.} {\bf B114} (1982) 431.

\bibitem{Ellis:1982ws}
J.~R. Ellis, D.~V. Nanopoulos, K.~A. Olive, and K.~Tamvakis, {\it {PRIMORDIAL
  SUPERSYMMETRIC INFLATION}},  {\em Nucl.Phys.} {\bf B221} (1983) 524.

\bibitem{Ellis:1982dg}
J.~R. Ellis, D.~V. Nanopoulos, K.~A. Olive, and K.~Tamvakis, {\it {Fluctuations
  in a Supersymmetric Inflationary Universe}},  {\em Phys.Lett.} {\bf B120}
  (1983) 331.

\bibitem{Langbein:1993ym}
R.~Langbein, K.~Langfeld, H.~Reinhardt, and L.~von Smekal, {\it {Natural slow
  roll inflation}},  {\em Mod.Phys.Lett.} {\bf A11} (1996) 631--646,
  [\href{http://arxiv.org/abs/hep-ph/9310335}{{\tt hep-ph/9310335}}].

\bibitem{GonzalezDiaz:1986bu}
P.~Gonzalez-Diaz, {\it {PRIMORDIAL KALUZA-KLEIN INFLATION}},  {\em Phys.Lett.}
  {\bf B176} (1986) 29--32.

\bibitem{Yokoyama:1998rw}
J.~Yokoyama, {\it {Chaotic new inflation and primordial spectrum of adiabatic
  fluctuations}},  {\em Phys.Rev.} {\bf D59} (1999) 107303.

\bibitem{Rehman:2008qs}
M.~U. Rehman, Q.~Shafi, and J.~R. Wickman, {\it {GUT Inflation and Proton Decay
  after WMAP5}},  {\em Phys.Rev.} {\bf D78} (2008) 123516,
  [\href{http://arxiv.org/abs/0810.3625}{{\tt arXiv:0810.3625}}].

\bibitem{Barenboim:2013wra}
G.~Barenboim, E.~J. Chun, and H.~M. Lee, {\it {Coleman-Weinberg Inflation in
  light of Planck}},  {\em Phys.Lett.} {\bf B730} (2014) 81--88,
  [\href{http://arxiv.org/abs/1309.1695}{{\tt arXiv:1309.1695}}].

\bibitem{Okada:2013vxa}
N.~Okada and Q.~Shafi, {\it {Observable Gravity Waves From $U(1)_{B-L}$ Higgs
  and Coleman-Weinberg Inflation}},  \href{http://arxiv.org/abs/1311.0921}{{\tt
  arXiv:1311.0921}}.

\bibitem{Hempfling:1996ht}
R.~Hempfling, {\it {The Next-to-minimal Coleman-Weinberg model}},  {\em Phys.
  Lett.} {\bf B379} (1996) 153--158,
  [\href{http://arxiv.org/abs/hep-ph/9604278}{{\tt hep-ph/9604278}}].

\bibitem{Okada:2011en}
N.~Okada, M.~U. Rehman, and Q.~Shafi, {\it {Non-Minimal B-L Inflation with
  Observable Gravity Waves}},  {\em Phys. Lett.} {\bf B701} (2011) 520--525,
  [\href{http://arxiv.org/abs/1102.4747}{{\tt arXiv:1102.4747}}].

\bibitem{Panotopoulos:2014hwa}
G.~Panotopoulos, {\it {Nonminimal GUT inflation after Planck results}},  {\em
  Phys. Rev.} {\bf D89} (2014), no.~4 047301,
  [\href{http://arxiv.org/abs/1403.0931}{{\tt arXiv:1403.0931}}].

\bibitem{Okada:2014lxa}
N.~Okada, V.~N. {\c S}eno{\u g}uz, and Q.~Shafi, {\it {The Observational Status
  of Simple Inflationary Models: an Update}},  {\em Turk. J. Phys.} {\bf 40}
  (2016) 150--162, [\href{http://arxiv.org/abs/1403.6403}{{\tt
  arXiv:1403.6403}}].

\bibitem{Marzola:2015xbh}
L.~Marzola, A.~Racioppi, M.~Raidal, F.~R. Urban, and H.~Veermäe, {\it
  {Non-minimal CW inflation, electroweak symmetry breaking and the 750 GeV
  anomaly}},  {\em JHEP} {\bf 03} (2016) 190,
  [\href{http://arxiv.org/abs/1512.09136}{{\tt arXiv:1512.09136}}].

\bibitem{Cerioni:2009kn}
A.~Cerioni, F.~Finelli, A.~Tronconi, and G.~Venturi, {\it {Inflation and
  Reheating in Induced Gravity}},  {\em Phys. Lett.} {\bf B681} (2009)
  383--386, [\href{http://arxiv.org/abs/0906.1902}{{\tt arXiv:0906.1902}}].

\bibitem{Rinaldi:2015yoa}
M.~Rinaldi, L.~Vanzo, S.~Zerbini, and G.~Venturi, {\it {Inflationary
  quasiscale-invariant attractors}},  {\em Phys. Rev.} {\bf D93} (2016) 024040,
  [\href{http://arxiv.org/abs/1505.03386}{{\tt arXiv:1505.03386}}].

\bibitem{Barrie:2016rnv}
N.~D. Barrie, A.~Kobakhidze, and S.~Liang, {\it {Natural Inflation with Hidden
  Scale Invariance}},  {\em Phys. Lett.} {\bf B756} (2016) 390--393,
  [\href{http://arxiv.org/abs/1602.04901}{{\tt arXiv:1602.04901}}].

\bibitem{Croon:2015fza}
D.~Croon, V.~Sanz, and J.~Setford, {\it {Goldstone Inflation}},  {\em JHEP}
  {\bf 10} (2015) 020, [\href{http://arxiv.org/abs/1503.08097}{{\tt
  arXiv:1503.08097}}].

\bibitem{Croon:2015naa}
D.~Croon, V.~Sanz, and E.~R.~M. Tarrant, {\it {Reheating with a Composite
  Higgs}},  \href{http://arxiv.org/abs/1507.04653}{{\tt arXiv:1507.04653}}.

\bibitem{Guth:2007ng}
A.~H. Guth, {\it {Eternal inflation and its implications}},  {\em J.Phys.} {\bf
  A40} (2007) 6811--6826, [\href{http://arxiv.org/abs/hep-th/0702178}{{\tt
  hep-th/0702178}}].

\bibitem{Starobinsky:1980te}
A.~A. Starobinsky, {\it {A New Type of Isotropic Cosmological Models Without
  Singularity}},  {\em Phys. Lett.} {\bf B91} (1980) 99--102.

\bibitem{Garny:2015sjg}
M.~Garny, M.~Sandora, and M.~S. Sloth, {\it {Planckian Interacting Massive
  Particles as Dark Matter}},  {\em Phys. Rev. Lett.} {\bf 116} (2016), no.~10
  101302, [\href{http://arxiv.org/abs/1511.03278}{{\tt arXiv:1511.03278}}].

\bibitem{Kofman:1994rk}
L.~Kofman, A.~D. Linde, and A.~A. Starobinsky, {\it {Reheating after
  inflation}},  {\em Phys. Rev. Lett.} {\bf 73} (1994) 3195--3198,
  [\href{http://arxiv.org/abs/hep-th/9405187}{{\tt hep-th/9405187}}].

\bibitem{Khlebnikov:1996wr}
S.~{\relax Yu}. Khlebnikov and I.~I. Tkachev, {\it {The Universe after
  inflation: The Wide resonance case}},  {\em Phys. Lett.} {\bf B390} (1997)
  80--86, [\href{http://arxiv.org/abs/hep-ph/9608458}{{\tt hep-ph/9608458}}].

\bibitem{Khlebnikov:1996zt}
S.~{\relax Yu}. Khlebnikov and I.~I. Tkachev, {\it {Resonant decay of Bose
  condensates}},  {\em Phys. Rev. Lett.} {\bf 79} (1997) 1607--1610,
  [\href{http://arxiv.org/abs/hep-ph/9610477}{{\tt hep-ph/9610477}}].

\bibitem{Prokopec:1996rr}
T.~Prokopec and T.~G. Roos, {\it {Lattice study of classical inflaton decay}},
  {\em Phys. Rev.} {\bf D55} (1997) 3768--3775,
  [\href{http://arxiv.org/abs/hep-ph/9610400}{{\tt hep-ph/9610400}}].

\bibitem{Greene:1997ge}
B.~R. Greene, T.~Prokopec, and T.~G. Roos, {\it {Inflaton decay and heavy
  particle production with negative coupling}},  {\em Phys. Rev.} {\bf D56}
  (1997) 6484--6507, [\href{http://arxiv.org/abs/hep-ph/9705357}{{\tt
  hep-ph/9705357}}].

\bibitem{Chung:1998bt}
D.~J.~H. Chung, {\it {Classical inflation field induced creation of superheavy
  dark matter}},  {\em Phys. Rev.} {\bf D67} (2003) 083514,
  [\href{http://arxiv.org/abs/hep-ph/9809489}{{\tt hep-ph/9809489}}].

\bibitem{Giudice:1999fb}
G.~F. Giudice, M.~Peloso, A.~Riotto, and I.~Tkachev, {\it {Production of
  massive fermions at preheating and leptogenesis}},  {\em JHEP} {\bf 08}
  (1999) 014, [\href{http://arxiv.org/abs/hep-ph/9905242}{{\tt
  hep-ph/9905242}}].

\bibitem{Farzinnia:2015fka}
A.~Farzinnia and S.~Kouwn, {\it {Classically scale invariant inflation,
  supermassive WIMPs, and adimensional gravity}},  {\em Phys. Rev.} {\bf D93}
  (2016), no.~6 063528, [\href{http://arxiv.org/abs/1512.05890}{{\tt
  arXiv:1512.05890}}].

\bibitem{Ade:2015ava}
{\bf Planck} Collaboration, P.~Ade et~al., {\it {Planck 2015 results. XVII.
  Constraints on primordial non-Gaussianity}},
  \href{http://arxiv.org/abs/1502.01592}{{\tt arXiv:1502.01592}}.

\bibitem{Chung:2004nh}
D.~J. Chung, E.~W. Kolb, A.~Riotto, and L.~Senatore, {\it {Isocurvature
  constraints on gravitationally produced superheavy dark matter}},  {\em
  Phys.Rev.} {\bf D72} (2005) 023511,
  [\href{http://arxiv.org/abs/astro-ph/0411468}{{\tt astro-ph/0411468}}].

\bibitem{inprogress}
L.~Marzola and F.~R. Urban, {\it {Ultra High Energy Cosmic Rays \& Super-heavy
  Dark Matter}},  \href{http://arxiv.org/abs/1611.07180}{{\tt
  arXiv:1611.07180}}.

\bibitem{Carbone:2008iz}
C.~Carbone, L.~Verde, and S.~Matarrese, {\it {Non-Gaussian halo bias and future
  galaxy surveys}},  {\em Astrophys.J.} {\bf 684} (2008) L1--L4,
  [\href{http://arxiv.org/abs/0806.1950}{{\tt arXiv:0806.1950}}].

\bibitem{Verde:2009hy}
L.~Verde and S.~Matarrese, {\it {Detectability of the effect of Inflationary
  non-Gaussianity on halo bias}},  {\em Astrophys.J.} {\bf 706} (2009)
  L91--L95, [\href{http://arxiv.org/abs/0909.3224}{{\tt arXiv:0909.3224}}].

\bibitem{Giannantonio:2011ya}
T.~Giannantonio, C.~Porciani, J.~Carron, A.~Amara, and A.~Pillepich, {\it
  {Constraining primordial non-Gaussianity with future galaxy surveys}},  {\em
  Mon.Not.Roy.Astron.Soc.} {\bf 422} (2012) 2854--2877,
  [\href{http://arxiv.org/abs/1109.0958}{{\tt arXiv:1109.0958}}].

\end{thebibliography}\endgroup

\end{document}